\documentstyle[twoside,psfig]{article}

\catcode`\@=11
\long\def\@makefntext#1{
\protect\noindent \hbox to 3.2pt {\hskip-.9pt  
$^{{\eightrm\@thefnmark}}$\hfil}#1\hfill}		

\def\thefootnote{\fnsymbol{footnote}}
\def\@makefnmark{\hbox to 0pt{$^{\@thefnmark}$\hss}}	
	
\def\ps@myheadings{\let\@mkboth\@gobbletwo
\def\@oddhead{\hbox{}
\rightmark\hfil\eightrm\thepage}   
\def\@oddfoot{}\def\@evenhead{\eightrm\thepage\hfil
\leftmark\hbox{}}\def\@evenfoot{}
\def\sectionmark##1{}\def\subsectionmark##1{}}

\oddsidemargin=\evensidemargin
\renewcommand{\thefootnote}{\fnsymbol{footnote}}

\newcounter{sectionc}\newcounter{subsectionc}\newcounter{subsubsectionc}
\renewcommand{\section}[1] {\vspace{12pt}\addtocounter{sectionc}{1} 
\setcounter{subsectionc}{0}\setcounter{subsubsectionc}{0}\noindent 
	{\tenbf\thesectionc. #1}\par\vspace{5pt}}
\renewcommand{\subsection}[1] {\vspace{12pt}\addtocounter{subsectionc}{1} 
	\setcounter{subsubsectionc}{0}\noindent 
	{\bf\thesectionc.\thesubsectionc. {\kern1pt \bfit #1}}\par\vspace{5pt}}
\renewcommand{\subsubsection}[1] {\vspace{12pt}\addtocounter{subsubsectionc}{1}
	\noindent{\tenrm\thesectionc.\thesubsectionc.\thesubsubsectionc.
	{\kern1pt \tenit #1}}\par\vspace{5pt}}
\newcommand{\nonumsection}[1] {\vspace{12pt}\noindent{\tenbf #1}
	\par\vspace{5pt}}

\newcounter{appendixc}
\newcounter{subappendixc}[appendixc]
\newcounter{subsubappendixc}[subappendixc]
\renewcommand{\thesubappendixc}{\Alph{appendixc}.\arabic{subappendixc}}
\renewcommand{\thesubsubappendixc}
	{\Alph{appendixc}.\arabic{subappendixc}.\arabic{subsubappendixc}}

\renewcommand{\appendix}[1] {\vspace{12pt}
        \refstepcounter{appendixc}
        \setcounter{figure}{0}
        \setcounter{table}{0}
        \setcounter{lemma}{0}
        \setcounter{theorem}{0}
        \setcounter{corollary}{0}
        \setcounter{definition}{0}
        \setcounter{equation}{0}
        \renewcommand{\thefigure}{\Alph{appendixc}.\arabic{figure}}
        \renewcommand{\thetable}{\Alph{appendixc}.\arabic{table}}
        \renewcommand{\theappendixc}{\Alph{appendixc}}
        \renewcommand{\thelemma}{\Alph{appendixc}.\arabic{lemma}}
        \renewcommand{\thetheorem}{\Alph{appendixc}.\arabic{theorem}}
        \renewcommand{\thedefinition}{\Alph{appendixc}.\arabic{definition}}
        \renewcommand{\thecorollary}{\Alph{appendixc}.\arabic{corollary}}
        \renewcommand{\theequation}{\Alph{appendixc}.\arabic{equation}}
        \noindent{\tenbf Appendix \theappendixc #1}\par\vspace{5pt}}
\newcommand{\subappendix}[1] {\vspace{12pt}
        \refstepcounter{subappendixc}
        \noindent{\bf Appendix \thesubappendixc. {\kern1pt \bfit #1}}
	\par\vspace{5pt}}
\newcommand{\subsubappendix}[1] {\vspace{12pt}
        \refstepcounter{subsubappendixc}
        \noindent{\rm Appendix \thesubsubappendixc. {\kern1pt \tenit #1}}
	\par\vspace{5pt}}

\newcommand{\textlineskip}{\baselineskip=13pt}
\newcommand{\smalllineskip}{\baselineskip=10pt}

\def\eightcirc{
\begin{picture}(0,0)
\put(4.4,1.8){\circle{6.5}}
\end{picture}}
\def\eightcopyright{\eightcirc\kern2.7pt\hbox{\eightrm c}} 

\newcommand{\copyrightheading}[1]
	{\vspace*{-2.5cm}\smalllineskip{\flushleft
	{\footnotesize International Journal of Modern Physics A #1}\\
	{\footnotesize $\eightcopyright$\, World Scientific Publishing
	 Company}\\
	 }}

\def\abstracts#1#2#3{{
	\centering{\begin{minipage}{4.5in}\footnotesize\baselineskip=10pt
	\parindent=0pt #1\par 
	\parindent=15pt #2\par
	\parindent=15pt #3
	\end{minipage}}\par}} 

\newcommand{\bibit}{\nineit}

\renewenvironment{thebibliography}[1]
	{\frenchspacing
	 \ninerm\baselineskip=11pt
	 \begin{list}{\arabic{enumi}.}
	{\usecounter{enumi}\setlength{\parsep}{0pt}
	 \setlength{\leftmargin 12.7pt}{\rightmargin 0pt} 
	 \setlength{\itemsep}{0pt} \settowidth
	{\labelwidth}{#1.}\sloppy}}{\end{list}}

\newcounter{itemlistc}
\newcounter{romanlistc}
\newcounter{alphlistc}
\newcounter{arabiclistc}

\newcommand{\fcaption}[1]{
        \refstepcounter{figure}
        \setbox\@tempboxa = \hbox{\footnotesize Fig.~\thefigure. #1}
        \ifdim \wd\@tempboxa > 5in
           {\begin{center}
        \parbox{5in}{\footnotesize\smalllineskip Fig.~\thefigure. #1}
            \end{center}}
        \else
             {\begin{center}
             {\footnotesize Fig.~\thefigure. #1}
              \end{center}}
        \fi}

\newcommand{\tcaption}[1]{
        \refstepcounter{table}
        \setbox\@tempboxa = \hbox{\footnotesize Table~\thetable. #1}
        \ifdim \wd\@tempboxa > 5in
           {\begin{center}
        \parbox{5in}{\footnotesize\smalllineskip Table~\thetable. #1}
            \end{center}}
        \else
             {\begin{center}
             {\footnotesize Table~\thetable. #1}
              \end{center}}
        \fi}

\def\@citex[#1]#2{\if@filesw\immediate\write\@auxout
	{\string\citation{#2}}\fi
\def\@citea{}\@cite{\@for\@citeb:=#2\do
	{\@citea\def\@citea{,}\@ifundefined
	{b@\@citeb}{{\bf ?}\@warning
	{Citation `\@citeb' on page \thepage \space undefined}}
	{\csname b@\@citeb\endcsname}}}{#1}}

\newif\if@cghi
\def\cite{\@cghitrue\@ifnextchar [{\@tempswatrue
	\@citex}{\@tempswafalse\@citex[]}}
\def\citelow{\@cghifalse\@ifnextchar [{\@tempswatrue
	\@citex}{\@tempswafalse\@citex[]}}
\def\@cite#1#2{{$\null^{#1}$\if@tempswa\typeout
	{IJCGA warning: optional citation argument 
	ignored: `#2'} \fi}}

\def\pmb#1{\setbox0=\hbox{#1}
	\kern-.025em\copy0\kern-\wd0
	\kern.05em\copy0\kern-\wd0
	\kern-.025em\raise.0433em\box0}

\def\fnt#1#2{\footnotetext{\kern-.3em
	{$^{\mbox{\scriptsize #1}}$}{#2}}}

\def\thefootnote{\fnsymbol{footnote}}
\def\@makefnmark{\hbox to 0pt{$^{\@thefnmark}$\hss}}	
	
\def\ps@myheadings{%
    \let\@oddfoot\@empty\let\@evenfoot\@empty
    \def\@evenhead{\slshape\leftmark\hfil}
    \def\@oddhead{\hfil{\slshape\rightmark}}
    \let\@mkboth\@gobbletwo
    \let\sectionmark\@gobble
    \let\subsectionmark\@gobble
    }
\font\tenrm=cmr10
\font\tenit=cmti10 
\font\tenbf=cmbx10
\font\bfit=cmbxti10 at 10pt
\font\ninerm=cmr9
\font\nineit=cmti9

\font\eightrm=cmr8

\def\qed{\hbox{${\vcenter{\vbox{			
   \hrule height 0.4pt\hbox{\vrule width 0.4pt height 6pt
   \kern5pt\vrule width 0.4pt}\hrule height 0.4pt}}}$}}

\renewcommand{\thefootnote}{\fnsymbol{footnote}}  

\def\lagr{\hbox{$\cal L$}}
\def\ve{\varepsilon} 
\def\w{\omega}

\def\P{\Psi}

\def\pd{\partial}

\def\L{\Lambda}

\def\W{\Omega} 
\def\z{\zeta}

\def\<{\langle} 
\def\>{\rangle}

\def\Log{\hbox{ln}} 
 
\def\a{\alpha} 
\def\b{\beta} 
\def\g{\gamma}   
   
   \def\L{\Lambda} 
\def\s{\sigma}

\def\m{\mu}

\def\z{\zeta} 
\def\w{\omega} 
\def\tt{\theta}

\def\({\left(} 
\def\[{\left[} 
\def\){\right)} 
\def\]{\right]}

\def\be{\begin{equation}}
\def\ee{\end{equation}}

\begin{document}
\setlength{\textheight}{7.7truein}  

\thispagestyle{empty}

\markboth{\protect{\footnotesize\it Casimir Energy
for Confined Fields...}}{\protect{\footnotesize\it Igor O. Cherednikov}}

\normalsize\textlineskip

\setcounter{page}{1}

\copyrightheading{}		

\vspace*{0.88truein}

\centerline{\bf CASIMIR ENERGY OF CONFINED FIELDS:}
\vspace*{0.035truein}
\centerline{\bf A ROLE OF THE RG-INVARIANCE} \vspace*{0.37truein}
\centerline{\footnotesize IGOR O. CHEREDNIKOV\footnote{Email: igorch@thsun1.jinr.ru,
igorch@itpm.msu.su}} \baselineskip=12pt \centerline{\footnotesize\it
Bogoliubov Laboratory of Theoretical Physics} \baselineskip=10pt
\centerline{\footnotesize\it Joint Institute for Nuclear Research,}
\baselineskip=10pt \centerline{\footnotesize\it 141980 Dubna, Russia}
\baselineskip=10pt \centerline{\footnotesize\it and }
\baselineskip=10pt \centerline{\footnotesize\it Institute
for Theoretical Problems of Microphysics,  Moscow State University,} 
\baselineskip=10pt \centerline{\footnotesize\it 119899 Moscow, Russia}
\baselineskip=10pt \centerline{\footnotesize\it and} \baselineskip=10pt
\centerline{\footnotesize\it the Abdus Salam International Centre for
Theoretical Physics,} \baselineskip=10pt \centerline{\footnotesize\it 34100
Trieste, Italy}

\vspace*{10pt}


\abstracts{A role of the renormalization group invariance in calculations 
of the ground state energy for models with confined fermion fields is
discussed. The case of the (1+1)D MIT bag model with the massive fermions
is studied in detail.}{}{}


\vspace*{1pt}\textlineskip	
\section{Introduction}	
\vspace*{-0.5pt}
\noindent
The calculations of the Casimir energy for quantized fields under nontrivial
boundary conditions encounter usually a number of difficulties (for the
most recent review on the Casimir energy see the ref.[1]). A majority of them
are connected with ambiguities in results obtained by means of different
regularization and renormalization methods. Physically more interesting
problem is the dependence of the (renormalized) energy from an additional mass
parameter, which emerges unavoidably in any regularization scheme. For
example, in the widely used $\z$-function regularization, the mass $\m$ must
be introduced in order to restore the correct dimension of the sum:  \be E = -
\sum \w_n \to E_{reg} (\m, \ve) = -\m^{\ve} \sum \w^{1-\ve}_n \ . \ee Whatever
renormalization procedure one applies, the finite part of the energy would
contain a $\m$-dependent contribution (recently this problem was addressed
and studied from the RG point of view in ref. [2]). Of course, there are
several situations, for which this dependence is canceled due to some
geometrical, or other, features of the given configuration. However, it would
be very useful and interesting to investigate more general case. 

In the present paper we consider the renormalization of the Casimir
energy from the point of view of the convenient quantum field theory, and
assume that variations of the mass scale $\m$ must not yield any physical
consequences. This requirement naturally leads to a sort of the
renormalization group equation, the solution of which allows to conclude that
some of the parameters of the ``classical'' mass formula have to be considered
as running constants. This may be important, { \it e. g.}, in some 
phenomenological applications, such as the quark bag models, since it may
provide an incite into the relations between fundamental and effective
aspects of the investigations of the hadronic structure.

\setcounter{footnote}{0}
\renewcommand{\thefootnote}{\alph{footnote}}

\section{Confined Massless Fermion Field in a Spherical Cavity: the MIT Bag
Model} 
\noindent
Let us illustrate the general ideas of the method with the simple example.
Consider the free fermion field confined to the spherical volume of the radius
$R$ under the MIT-bag boundary conditions: 
\be
(i  n_\a \g^\a +1) \P (R) = 0 \ \ , \ \ n_\a = { r_\a \over r} \ . 
\ee
These conditions provide the absence of the quark flow through the surface of
the bag$^3$. 

As is well-known, the Casimir energy of such configuration defined as the
$\z$-regularized sum (1), where $\w_n$ are the eigenvalues of the free Dirac
Hamiltonian $H = -i \a \pd + \b m $, contains the terms singular in the limit
$\ve \to 0$, that are proportional to the powers of $R$ from $R^3$ to
$R^{-1}$. These singularities have to be absorbed into the definitions of
the corresponding constants in the ``classical'' mass formula$^4$ 
\be
E_{cl} = B {4 \over 3} \pi R^3 + \s 4 \pi R^2 + f R + \L + h R^{-1} \ . 
\ee
In the case of a massless field, the only divergent term survives that is
proportional to $R^{-1}$, and the renormalized total energy of the MIT bag
without valence quarks reads (for the sake of simplicity, we keep only the
volume part in the classical energy): 
\be
E_{MIT} = {1 \over R} (h - h_1 \Log \m R ) + {4 \over 3} \pi R^3 B \ , 
\ee
where $B$ is the bag constant, $h_1$ can be calculated by means of, { \it e.
g.}, heat-kernel technique (it had been shown$^4$ that $h_1 = {1
\over 63 \pi}$), and $\m$ is the additional arbitrary mass scale. 

The condition of the independence of $E_{MIT}$ from a choice of the value
of $\m$ leads to the functional equation
\be
\m {d  \over d\m} E_{MIT} = 0 \ ,  \label{rg}
\ee
what means that $h$ should be considered as a ``running constant'': 
\be
h(\m) = h_1 \Log {\m \over \m_0} \ , 
\ee
where $\m_0$ is the normalization point. Therefore, the bag energy equals 
\be
E_{MIT} (R, \m_0) = - {1 \over R } h_1 \Log \m_0 R + {4 \over 3} \pi R^3 B \ .
\ee

If we want to consider this bag as a model of a stable composite object, like
hadron, we need its energy to have a minimum at a certain value of the radius
$R$. So, the condition of the RG invariance for the energy (\ref{rg}) must be
supplied with the equation
\be
{d \over dR} E_{MIT} (R, \m_0) = 0 \ , \label{rad}
\ee
what yields 
\be
\Log \m_0 R = - \( {8 \over 3} {\pi R^4 B \over h_1} +1 \) \ . 
\ee
Equations (\ref{rg}) and (\ref{rad}) give the minimal radius $R_{min}$ as a
function of the energy scale 
\be
R_{min} = R_{min} (\m_0) \ . 
\ee
It's clearly seen that $R_{min}$ decreases when $\m_0$ increases, what
corresponds to the idea of the bag model as an effective approach to the
investigation of the hadronic structure at some low energy scale.

\section{(1+1)D MIT Bag Model with Massive Fermions}
\noindent
The presence of a mass may lead, in general,  to some new divergences that
have to be subtracted. Consider in detail the (1+1)D MIT bag model with the
massive fermions$^3$. The Lagrangian of this system
\be
\lagr_{MIT} = i\bar\psi\g\pd\psi - \bar\psi\psi\(m\ \tt(|x|<R) +
M\ \tt(|x|>R)\)  \ee
describes (in the limit $M\to\infty$) the fermion field confined to the segment $[-R,R]$
under the (1+1)D boundary condition: 
\be
(\pm i \g^1 +1)\psi(\pm R) = 0\ . 
\ee
The exact spectrum of the elementary fermionic excitations reads: 
\be
\w_n = \sqrt{\({\pi \over 2R} n + {\pi \over 4R}\)^2 + m^2} \ .
\ee
Here we will be interesting only in the small mass $m$ limit, so we drop out
all terms of the order $m^4$ and higher$^5$. Then the eigenvalues $\w_n$ can be
written as
\be
\w_n = \W_1 n + \W_0 + {m^2 \over 2\(\W_1 n +\W_0\)} + O(m^4) \ , 
\ee 
where
\be
\W_1 = {\pi \over 2R} \ \ , \ \ \W_0 = {\pi \over 4R} \ . 
\ee
In order to analyze the singularities in the Casimir energy, we use the
expansion for $n > 0$: 
\be
\w_n = \W_1 n + \W_0 + {\W_{-1} \over n} + O(n^{-2}) \ , 
\ee
where $\W_{-1} = m^2 R/\pi$, and assume the lowest valence state with $\w_0 =
\W_0 + 2\W_{-1}$ to be filled. 

It can be shown, that the $\z$-regularized sum (1) reads:
\be
E_{reg} = -\W_{-1} \({1 \over \ve} + \g_E \) + {\W_1 \over 12} + {\W_0 \over
2} + {\W_0^2 \over 2 \W_1} + \W_{-1}\(\Log {\W_1 \over \m} +1 \) , \label{div}
\ee where $\g_E = 0.5772...$ is the Euler constant. 
It's interesting to note, that the regularization by the exponential cutoff
gives the equivalent result$^5$. 

The divergent part of (\ref{div}) can be extracted in the form: 
\be
E_{div} = - {m^2 R \over \pi} \({1 \over \ve} + \g_E - \Log {\pi \over 8} \ .
\label{ms} \) 
\ee 
We include in $E_{div}$ the pole $\ve^{-1}$  as well as the
transcendent numbers $\g_E$ and $\Log {\pi \over 8}$ in analogy to the widely
used scheme {\sl MSbar} in QFT, but we should mention that this analogy is
only formal one, since (\ref{ms}) has
nothing to do with the singularities appearing in the conventional field
theory since it depends on the geometrical parameter $R$. This kind of
divergences containing the dependence from a dimensional parameter is close to
the so-called cusp singularity in the Wilson loops renormalization, and may be
treated on the same basis$^6$. 

The renormalization of (\ref{div}) is performed by the absorption of $E_{div}$
(\ref{ms}) into the definition of the ``classical'' bag constant $B$, which is
introduced in the mass formula and characterizes the energy excess inside the
bag volume as compared to the energy of nonperturbative vacuum outside$^3$. 

The finite renormalized energy of our bag with one valence fermion on the
lowest energy level is
\be
E(R, \m) = 2B_0 R + {11 \pi \over 48 R} + {3 m^2 R \over \pi} - {m^2 R \over
\pi } \Log \m R  \ , \ee where $B_0$ is the renormalized bag constant. 
Taking into account the RG equation (\ref{rg}), we find that $B_0$ should be
treated as the running constant
\be
B_0 (\m) = {m^2 \over 2 \pi} \Log {\m \over \m_0} \  
\ee
with the normalization point $\m_0$. Therefore, the total energy is 
\be
E_{MIT} = - {m^2 R \over \pi} \Log \m_0 R + {11 \pi \over 48 R} + {3
m^2 R \over \pi} \ . \ee The condition of stability (\ref{rad}) may be written
as \be
\Log \m_0 R = 2 - {11 \pi^2  \over 48 m^2 R^2} \ , 
\ee and hence we obtain the radius of a stable bag $R_{min}$ as a function of
the energy scale, decreasing with growth of $\m_0$. 

\section{Conclusion}
\noindent
We have analyzed the consequences of the condition of renormalization group
invariance in the Casimir energy calculations on the simple examples related
to the quark bag models. It is shown, that the requirement of the stability
supplied with the RG invariance leads to the dependence of the bag's size
$R_{min}$ on the mass renormalization point $\m_0$ which characterizes the
energy scale, and therefore, the validity of the bag approximation.

\nonumsection{Acknowledgements}
\noindent
I thank the Organizators of the QFEXT01 and especially Dr. Michael
Bordag for financial support and hospitality during this conference. I
acknowledge the discussions with Drs. M. Bordag, D. Vassilevich, and O.
Pavlovsky on the various aspects of my report. I thank Prof. I.
Brevik for drawing my attention to the paper [2]. This work is supported in
part by RFBR (projects 01-02-16431 and 00-15-96577), and INTAS (project
00-00-366). I'm grateful also to the Abdus Salam ICTP in Trieste where this
paper was completed.

\eject

\end{document}